\documentclass{emulateapj}
\newcommand\eq{\begin{equation}}
\newcommand\eeq{\end{equation}}
\newcommand\eqn{\begin{eqnarray}}
\newcommand\eeqn{\end{eqnarray}}

\newcommand\Msol{M$_{\odot}$}
\newcommand\kms{km s$^{-1}$}
\newcommand\ergs{erg s$^{-1}$}

\newcommand\sloani{$i^\prime$}
\newcommand\sloang{$g^\prime$}
\newcommand\RXJ{RX~J1416.4+2315$\,$}
\newcommand\RXJold{RX~J1552.2+2013$\,$}

\shorttitle{The compact group -- fossil group connection}
\shortauthors{Mendes de Oliveira \& Carrasco}
\slugcomment{ApJ Letter, accepted}

\begin{document}

\title{The compact group--fossil group connection: observations of a 
massive compact group at z$=$0.22\altaffilmark{1}}

\altaffiltext{1}{Based on observations obtained at the Gemini Observatory,
which is operated by the Association of Universities for Research in
Astronomy, Inc., under a cooperative agreement with the NSF on behalf of
the Gemini partnership: the National Science Foundation (United States),
the Particle Physics and Astronomy Research Council (United Kingdom),
the National Research Council (Canada), CONICYT (Chile), the Australian
Research Council (Australia), CNPq (Brazil) and CONICET (Argentina) --
Program ID: GS-2005B-Q-37}

\author{Claudia L. Mendes de Oliveira}
\affil{Departamento de Astronomia, Instituto de Astronomia,
Geof\'isica e Ci\^encias Atmosf\'ericas, Universidade de S\~ao Paulo,
Rua do Mat\~ao 1226, Cidade Universitaria, 05508--090,
S\~ao Paulo, Brazil; oliveira@astro.iag.usp.br}
\and
\author{Eleazar R. Carrasco}
\affil{Gemini Observatory, Southern Operations Center, AURA, Casilla 603,
La Serena, Chile; rcarrasco@gemini.edu}

\begin{abstract}
It has been suggested that fossil groups could be the canibalized remains
of compact groups, that lost energy through tidal friction. However,
in the nearby universe,  compact groups which are close to the merging
phase and display a wealth of interacting features (such as HCG 31 and
HCG 79) have very low velocity dispersions and poor neighborhoods,
unlike the massive, cluster-like fossil groups studied to date. In
fact, known z$=$0 compact groups are very seldom embedded  in massive
enough structures which may have resembled the intergalactic medium of
fossil groups.   In this paper we study the dynamical properties of CG6, a
massive compact group at z$=$0.220 that has several properties
in common with known fossil groups.  We report on new \sloang~and
\sloani~imaging and multi-slit spectroscopic performed with
GMOS on Gemini South. The system has 20  members, within a radius of 1 
h$_{70}^{-1}$ Mpc, a velocity dispersion of 700~\kms~and has a mass of 
1.8 $\times$ 10$^{14}$ h$_{70}^{-1}$~\Msol, similar to that of the most 
massive fossil groups known. The merging of the four central galaxies in 
this group would form a galaxy with magnitude $M_{r'} \sim -23.4$, typical 
for first-ranked galaxies of fossil groups. Although nearby compact groups 
with similar properties to CG 6 are rare, we speculate that such systems 
occurred more frequently in the past and they may have been the precursors 
of fossil groups.  
\end{abstract}


\section{Introduction}\label{sec:intro}

Groups of galaxies are small systems of typically a few L$^{*}$ galaxies,
which comprise over 55\% of the nearby structures in the universe. A
small fraction of galaxy groups are classified as {\it compact}
groups, which are responsible for $\sim$ 1\% of the luminosity
density of the universe \citep{mdoh91}. Although they are rare
objects in the nearby universe, their high galactic densities and low
velocity dispersions make them ideal systems for the study of galaxy
transformation through galaxy-galaxy collisions.  As expected, these
systems have a high fraction of interacting members, although merged
objects are rare \citep{zepf93}.  They are commonly believed to
evolve through dynamical friction and finally merge to form one
single galaxy \citep{Barnes92}. \citet{vikhlinin99} and
\citet{jones03} have suggested that the merging of compact groups 
can lead to the formation of  fossil groups.  
A fossil group (FG, hereafter) is a
system with an extended and luminous X-ray halo (L$_X$ $>$ $10^{42}$
h$_{70}^{-2}$ erg s$^{-1}$), dominated by one single brighter than
L$^{*}$ elliptical galaxy, surrounded by low-luminosity companions 
\citep[where the difference in magnitude between the bright dominant elliptical and the next  brightest companion is $>$ 2 mag in the R-band;][]{jones03}.

One important goal of this article is to investigate if compact groups (CG, 
hereafter)
as we known them today, could be the precursors of FGs.
In order to answer this question, we summarize, in section 2, the
properties of a few of the most strongly interacting nearby CGs
known, which are about to merge, and in section 3, we describe the
properties of the five FGs which have been studied
spectroscopically so far. In section 4, we present new observations
for a CG embedded in a cluster-size potential, 
at redshift z$=$0.22 and section 5 puts together all the observations
described and discusses the CG-FG scenario. Throughout this paper we adopt
when necessary a standard cosmological model:
$H_{0}=70\,h_{70}\,$km~s$^{-1}$~Mpc$^{-1}$, $\Omega_{m}=0.3$ and
$\Omega_{\Lambda}=0.7$. At z$=$0.22, 1\arcsec~corresponds to $ 3.5\,
h_{70}^{-1}\,$kpc.

\section {Interacting compact groups: HCG 31, HCG 79 and HCG 92}
\label{sec:hcg}

  There is evidence from both observations and simulations
that groups evolve through dynamical friction and coalesce to
form more compact structures as the universe ages. A few of
the most compact, and therefore most evolved groups known,
from Hickson's catalogue \citep{h92}  are HCG 31, HCG 79 (or
Seyfert Sextet) and HCG 92 (or Stephan's quintet).  The study
of these groups is important to help understanding
processes common in merging systems, environments that may
have occurred more often in the high-redshift universe.

HCG 31 is a group at z$\sim$0.013 and with a velocity dispersion of
$\sigma$ $\sim$ 60 \kms. This is a gas-rich group with intense star
forming activity \cite[e.g. ][]{moetal06, amram07}, dominated by a central
pair of interacting dwarf galaxies A$+$C. HCG 31 is thought to be in a
pre-merging phase \citep{amram05,vm05} and it has well developed tidal
tails seen in H$\alpha$ and HI. The group hosts two excellent candidates
for tidal dwarf galaxies, namely member F, in the southern tail and
member R, 50 h$_{70}^{-1}$ kpc to the north of the group (for an assumed 
distance modulus of DM=33.8).

  HCG 79, also known as {\it Seyfert Sextet}, was originally
identified as a sextet of galaxies but it is now known to be a
quartet at z$=$ 0.0145 (the 5th object is in
the background and the 6th is a luminous tidal debris to the
northeast of the group).  This is the most CG in
Hickson's catalogue with a galaxy-galaxy distance below
10 kpc (for an adopted DM=34.0)
and a velocity dispersion of $\sigma$ = 138 \kms.   
The four galaxies present morphological
distortions and increased activity (tidal debris, bar in HCG
79B, dust lane in HCG 79A, radio and infrared emission,
disturbed rotation curves and nuclear activity).  The group
presents a prominent intra-group light envelope which contains
45\% of the total light of the group \citep{darocha05}  and
irregular envelopes of HI \citep{wil91} and X--rays
\citep{pil95}. These suggest that recent or on-going
interaction is taking place within this system.

   HCG 92, also known as Stephan's quintet, is in reality a  quartet
with z$=$ 0.0215  and a foreground galaxy. It is the most well
studied  CG -- multi wavelength data are available from
radio to X-rays. Most of the gaseous material in Stephan's quintet is
concentrated not in the galaxies but in the intragroup
medium, suggesting that collisions among group members may have
happened frequently. A number of tidal dwarf galaxies and intergalactic
HII regions have been
identified in this group \citep[e.g.][]{moetal01,moetal04,xu05}.
Of the three groups
described above, HCG 92 is the only one to have detected X-rays, with
a total bolometric luminosity of  2.96$\times$10$^{42}$ h$^{-2}_{70}$
\ergs~\citep{Xue2000}. 

These three spiral-rich groups are thought to be in their
final stages of evolution -- they are, in fact, some of the
most compact systems found in the Hickson's catalogue. Yet,
they have members that can be clearly identified as individual
galaxies, suggesting that once merging starts, it may proceed
quickly, and the groups may no longer be recognized as such. 
The bright members of these  groups will almost certainly end
up as a single galaxy pile. A discussion of 
whether these systems will most likely end up as
FGs or as single isolated elliptical galaxies is deferred
to section 5. In the following section, some of the optical 
properties of the FGs studied so far are summarized.

\section{Dynamical properties of FGs} \label{sec:fossil}

Only five FGs have been studied so far in any level of
detail in the
optical bands (imaging and spectroscopy). 

\citet{mo06} derived the physical properties of the FG \RXJold,
at a redshift of z$=$0.136, and computed its luminosity function, based
on the spectroscopy of 36 member galaxies.  This system was found to
be a fossil cluster, given its high number of members and high velocity
dispersion (close to 700 km/s).

\cite{cypriano06} studied a second FG, \RXJ, at a
similar redshift of z=0.137. For this system also a fairly
high velocity dispersion was measured (584 km s$^{-1}$), for
25 members located in the inner 542 h$_{70}^{-1}$ kpc ($\sim$ 0.45 
of the virial radius) of the system. Similar results were found by
\citet{habib06}.

In two recent studies Mendes de Oliveira et al. (2007, in preparation)
and Cypriano et al. (2007, in preparation) two other
FGs were also  found to have cluster-like masses: RX
J1340.5+4017,  with 25 members, was found to have a velocity
dispersion of 580~\kms\ and for  RX J1256.0+2556, a velocity
dispersion of 582~\kms\ was determined from spectroscopy of 28
members. In all four cases the systems presented very pronounced
red  sequences in their color-magnitude diagrams and they were
dominated by early-type galaxies. Table 3 of \citet{habib06} lists
one other FG for which spectroscopy for a significant
number of members has been  obtained, ESO 3060170, which is found to
have a velocity dispersion of 648~\kms\, derived from velocity
measurements of
15 members. 

  The conclusion is then that for all five groups studied so far, 
for which more than six members are known, they all have 
velocity dispersions of $\sim$ 600~\kms\ or higher, and dynamical
virial masses $\sim$ 10$^{14}$ h$_{70}^{-1}$ \Msol.

  Fossil groups were suggested to be the end products of
merging of L$^{*}$ galaxies in low-density environments
\citep{jones03}.  However, the five FGs studied
do not constitute low-density environments and, in
fact, are more similar to galaxy clusters. The fairly high
X-ray emission, the large fraction of elliptical galaxies,
as well as the lack of obvious substructures,
suggest that these FGs are fairly massive
virialized systems.

   The FGs described above have masses, kinematics  and
environments which are  very different from those of nearby CGs.
At first sight,  this evidence would seem to indicate that there
is no connection between these two kinds of systems. However,
before discussing this point any further, we would like to
present in the next section, an example of a CG that does have
similar optical properties to those of FGs. It is the farthest
away CG known to date, at z$=$0.22. New imaging and
spectroscopic observations of this  group are presented in the
next section. A detail description of the properties of 
CG 6 will be presented in a future paper (Carrasco et al.
2007, in preparation).  

\section{CG 6 - a compact group at z=0.22}\label{sec:imaobs}

CG 6 is a group from the \citet{lee04} catalog of CG 
galaxies, chosen from the SDSS database. Before the present study,
only one galaxy in the system, the central one, had a measured
redshift.  The observations of SDSS CG 6 were carried out using the
Gemini Multiobject Spectrograph \citep[hereafter GMOS][]{hook04} at 
the Gemini South telescope. 

\begin{figure}[!htb]
\centering
\includegraphics[width=0.9 \columnwidth]{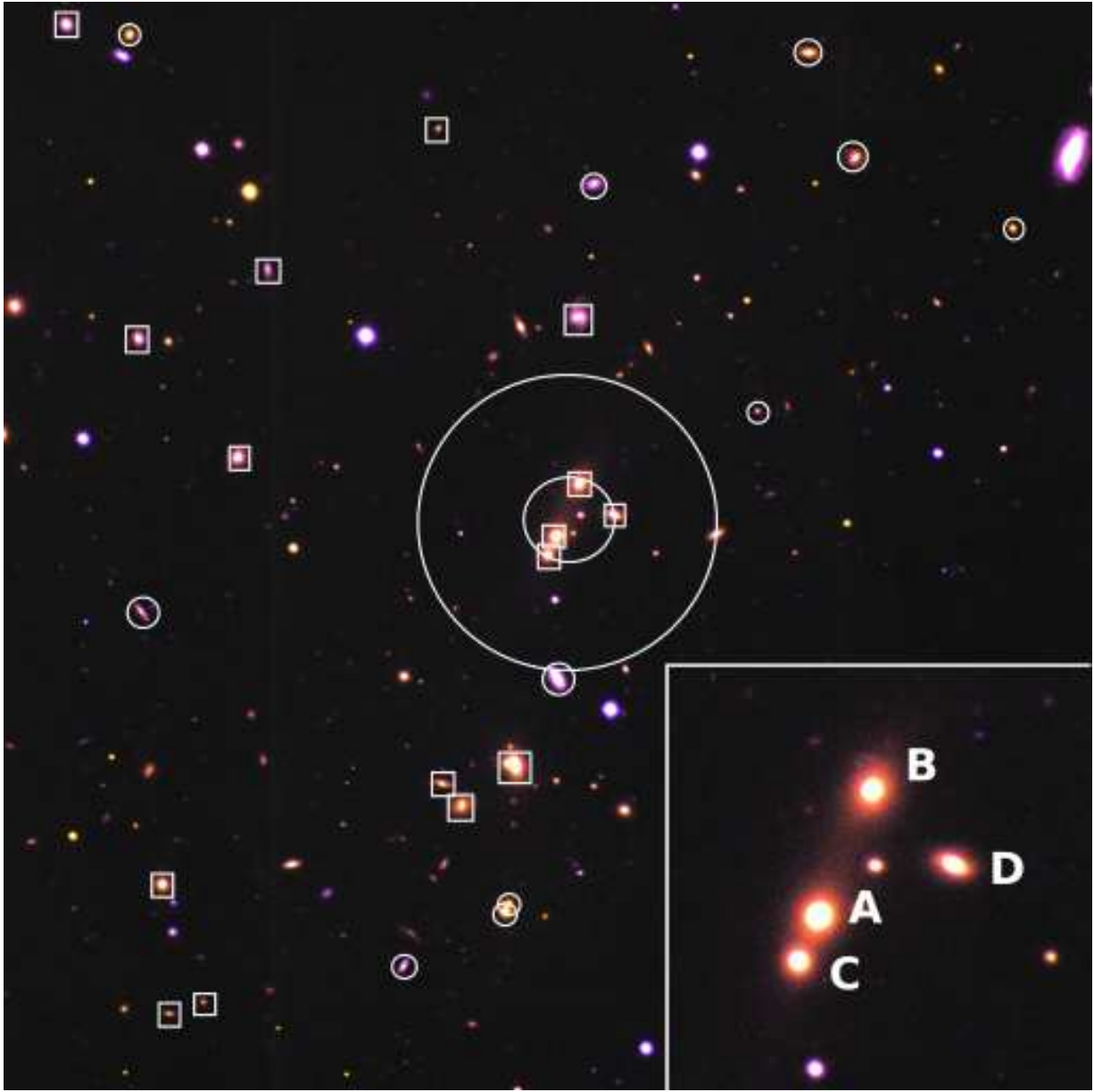}%
\caption[]{Color composite image of SDSS-CG-6. The field of
view is 5\farcm15 on a side (1.12 h$^{-1}_{70}$ Mpc at the
distance of the group). The squares are galaxies members of
the group. The circles are galaxies in the foreground and
background of the group. The inner circle is the smallest circle 
that  contains the centers of the group members with magnitudes  
within 3 mag from the  brightest group member. The outer concentric 
circle corresponds to the angular diameter that contains no other 
(external) galaxies within 3 mag from the brightest group member, 
as defined by \citet{lee04}. The zoom (inner 0.2 h$^{-1}_{70}$ Mpc) 
shows the four central galaxies in the group.}
\label{sloanimage}
\end{figure}

The group was imaged using the standards \sloang~and \sloani~filters
\citep{fuk96} in December 1$^{st}$, 2005. Exposures of 3$\times$150 
seconds in \sloang~and 3$\times$120 seconds in \sloani~filter, under
photometric conditions, were obtained. The images were observed with 
median seeing values of 0$\farcs$73 and 0$\farcs$55 in \sloang~and 
\sloani~filters respectively. Figure \ref{sloanimage} shows the color
composite image of the SDSS-CG-6 group. All  galaxies with measured 
radial velocities are marked (see below).

All observations were processed with the Gemini IRAF package
version 1.8 inside IRAF\footnote[2]{IRAF is distributed by  NOAO,
which is operated by the Association  of Universities for
Research in Astronomy Inc., under contract with the National
Science Foundationand} in a standard way. Calibrations of the 
magnitudes to the standard system were derived using observations 
of standard stars from \citet{lan92} field RU149. 

Object detection and photometry were performed on the 
\sloani-band image with the SExtractor program \citep{ber96}. 
{\rm MAG$\_$AUTO} was adopted as the total magnitude. Colors were 
derived by measuring fluxes inside a fixed circular  aperture of 24 pixels 
(1$\farcs$75) in both filters, corresponding to a physical aperture 
of 6.2 h$^{-1}_{70}$ kpc at the rest frame of the group.
All objects with SExtractor {\em stellarity} flag $\le0.85$ were
selected as galaxies. We estimate that the  catalog is complete down 
to \sloani$=$23.5 mag, since the number counts start to turn over 
at this value. The final 
catologue contains the total magnitudes,  colors and structural  
parameters for 409 galaxies brighter than \sloani$=23.5$
($M_{i^{'}}\sim-16.7$ mag at the  distance of the group, with no
k-correction).

\begin{figure}[!htb]
\centering
\includegraphics[width=0.9 \columnwidth]{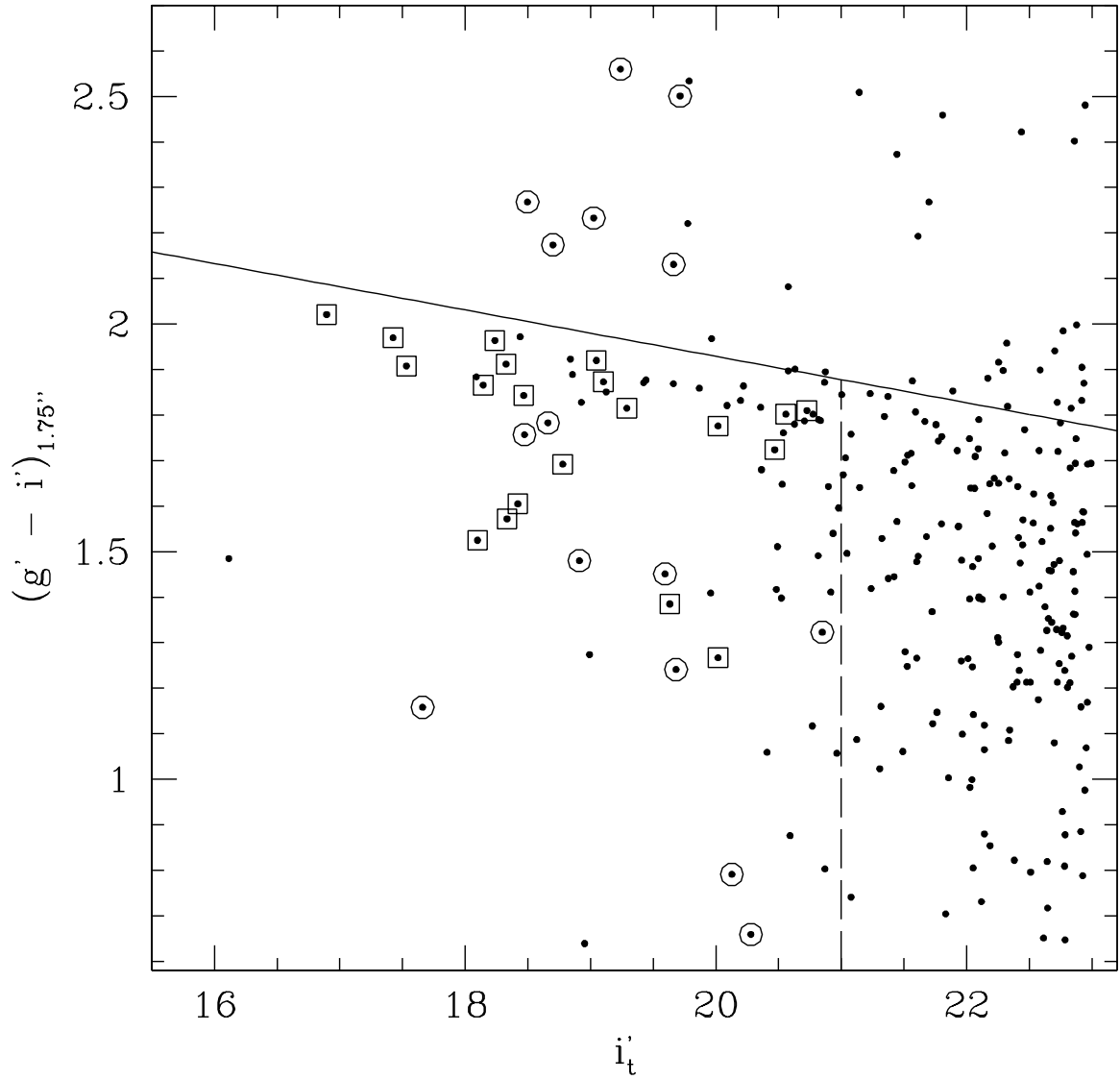}
\caption{Color-magnitude diagram of all galaxies brighter than
23 mag in the CG 6 field ({\em dots}). Colors and
magnitude are corrected for  galactic extinction. The symbols
are the same as in Figure~\ref{sloanimage}. The solid and dashed
lines indicate the upper limit for the red sequence and the
limiting magnitude we adopted for the spectroscopic target
selection.}
\label{colormag}
\end{figure}

Galaxies for spectroscopic follow-up were selected based on
their magnitudes and colors. Figure \ref{colormag} shows the
color-magnitude diagram for galaxies with \sloani$\le$23 mag. A 
pronounced red sequence is readly seen. Objects in the region 
below the red cluster sequence (continuous line) and brighter 
than \sloani$=21$ mag ($M_{i'}<-19.2$, (dashed line) were selected 
as potential candidates of the system. 

Galaxy spectra were observed in December 2005, during dark time,
with relatively good transparency and with a seeing that varied
between  0$\farcs$9 and 1$\farcs$0. Three exposures of 1400 seconds
were obtained through a mask with 1$\farcs$0 slits, using the 400 
lines mm$^{-1}$ ruling  density grating (R400) and centered at 
6300\AA, for a resolution of 8\AA. Reduction and calibration of
the data were done in the standard way, using the Gemini IRAF
package version 1.8.

\begin{figure}[!htb]
\centering
\includegraphics[width=0.9 \columnwidth]{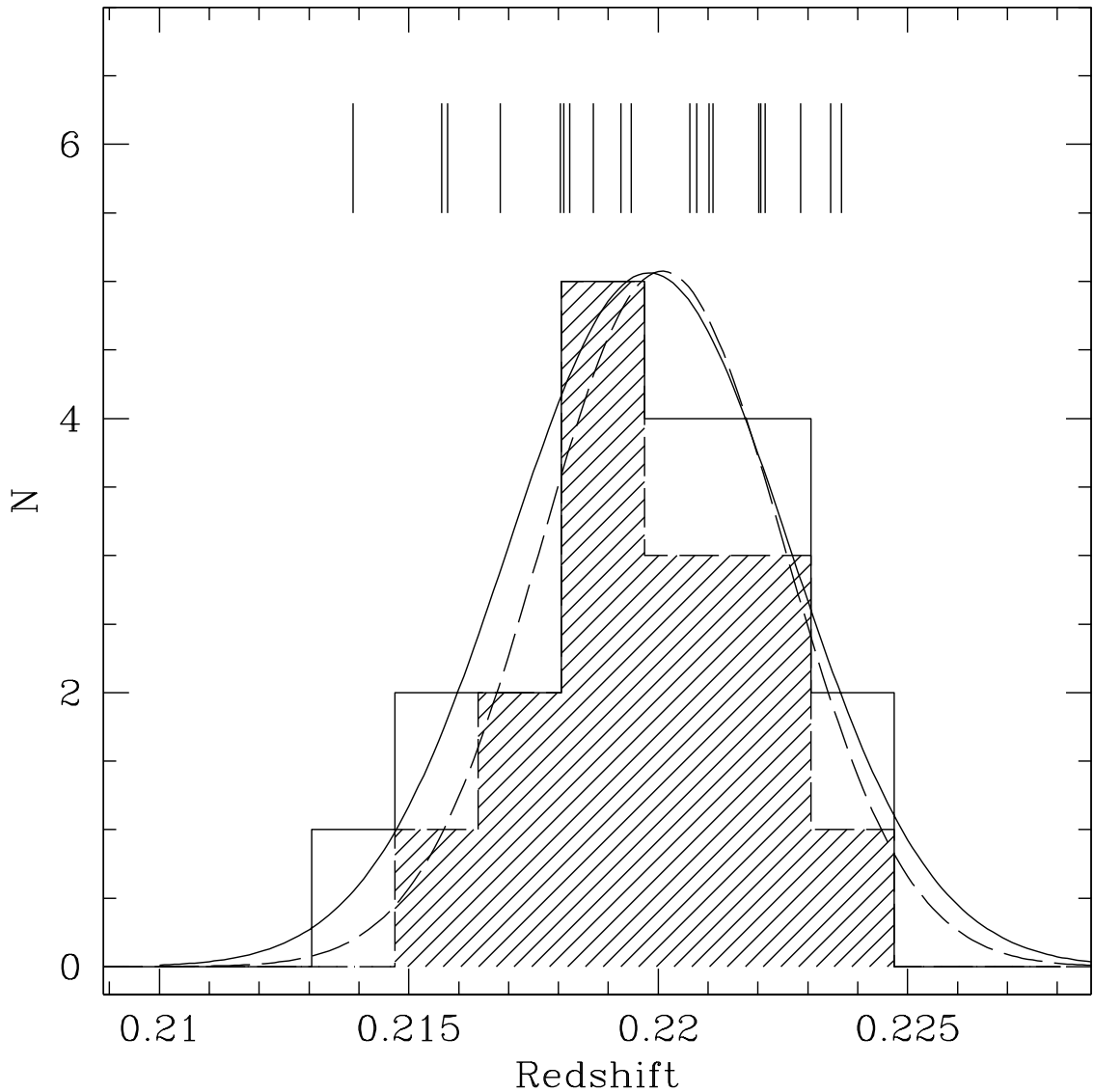}%
\caption{Histogram of the redshift distribution of 20 galaxies
in area of SDSS-CG-6. The average redshift and the velcity
dispersion for the whole sample (20 galaxies) is $\langle z
\rangle=0.21981$ and $\sigma=703$ km s$^{-1}$ (continuous
line). The hashed histogram show the velocity distribution for
the 15 non-emission line galaxies in the sample. For this
sub-sample, the average redshift is $\langle z
\rangle=0.21837$ and the velocity dispersion is $\sigma=608$
km s$^{-1}$ ({\em dashed line}).} 
\label{histovel}
\end{figure}

We were able to measure redshifts for 35 observed galaxies 
using cross-correlation techniques or emission-line fitting 
Twenty two of the galaxies are located at the redshift
of the group (within $\pm2000$ km s$^{-1}$ and between
0.213$<$z$<$0.226). We used the ROSTAT program \citep{bee90} 
to calculate the average velocity and the one-dimensional 
line-of-sight velocity dispersion of the group. We found 
that the group is located at $\langle z \rangle=0.21981\pm0.00054$, 
with a velocity dispersion of $\sigma=703\pm103$ km s$^{-1}$, and
with 20 member galaxies. Figure \ref{histovel} shows the
velocity distribution of the 20 galaxies at the redshift of
the group. We used the same software to do a statistical analysis 
of the velocities and we found no large gap in the velocity distribution, 
which follows close to a gaussian shape. The virial mass computed 
using the 20 galaxies with concordant redshifts, for a distance 
modulus of 40.18 is M$_{vir}=1.77_{-0.21}^{+0.45}$ $\times$ 10$^{14}$ h$^{-1}_{70}$
\Msol\ and the virial radius is R$_{vir}=0.82_{-0.01}^{+0.16}$ 
h$^{-1}_{70}$ Mpc (the errors are 68\% confidence intervals).

We checked for any significant effects in the velocity
dispersion and total mass of the system when we exclude the
emission line galaxies from the sample. The hashed histogram in
Figure \ref{histovel} shows the distribution of the 15 non-emission
line galaxies. For this smaller sample, we obtained a lower velocity
dispersion of $\sigma=608\pm99$ km s$^{-1}$, a virial mass of
M$_{vir}=0.95_{-0.11}^{+0.32}$ $\times$ 10$^{14}$ h$^{-1}_{70}$ 
\Msol\ and a virial radius R$_{vir}=0.64_{-0.02}^{+0.18}$ 
h$^{-1}_{70}$  Mpc (again the errors are 68\% confidence intervals),
values not significantly different from those considering the whole
sample. The results go in the direction of what was previously
seen also for other systems: inclusion of the emission-line galaxies
enhance the velocity dispersion of the system.

\section{Discussion}

Dynamical friction and subsequent merging are probably the processes
responsible for the lack of L$^{*}$ galaxies in FGs. 
Considering the merging scenario, it is possible that the
overluminous central galaxy in a FG has been formed within
a substructure, inside a larger structure.  In that case, one could
think of a scenario where a CG was formed within a rich
group, which would then have merged, leaving  behind the
brightest elliptical galaxy of what today is seen as a FG.
One weak argument against this scenario is that the nearby examples
of CGs are not usually found within such massive
structures, but instead are more often surrounded by very sparse
structures. There are, however, examples such as CG6, surrounded by
large numbers of lower-luminosity galaxies, inhabiting a deep potential
well.

We would like to test the hypothesis that CGs, as
observed in the nearby universe, could be the precursors of FGs.
We may examine two aspects: (1) if the sum
of the  luminosities of the brightest CG galaxies is
similar to the luminosity of a first-ranked FG galaxy and;
(2) if the  neighbourhoods of CGs are rich, i.e., if the
system as a whole  (group plus environment) has a velocity
dispersion/mass similar to that of a FG.

We compute the total luminosity of the galaxies in the soon-to-merge
CGs,  HCG 31, HCG 79 and HCG 92, to check how these compare with the
luminosities of first-ranked galaxies in FGs.  Adding up the luminosities
of galaxies HCG 31 A to C, G and Q, which are the brightest in the
group HCG 31, a magnitude of M$_R = -22.5$ is obtained (for a distance
modulus, DM = 33.8).  Summing up the luminosities of galaxies HCG 79 A-D,
an equal total magnitude of M$_R=-22.5$ is obtained (for DM = 34.0).
These are upper limits on the luminosities of these objects given that
several members are starburst galaxies. After fading, the merged central
object in HCG 31 and HCG 79 will have somewhat lower magnitudes than that
of a typical first-ranked galaxy in a FG.  Fossil groups first-ranked
galaxies have luminosities well above L$^{*}$. For the five FGs studied by
\citet{jones03}, the first-ranked galaxies had a median luminosity
of $M_R=-23.2$ and for the 34 FGs found in the SDSS DR5 by \citet{dosSantos07}
the median luminosity was $M_R=-23.5$.

Although for HCG 92, the final object (adding up luminosities of galaxies
A-E) would have an absolute magnitude of $M_R=-24.2$ (for DM$=$34.8), 
which after allowing for some fading, could be similar to that of
an FG first-ranked galaxy, HCG 92 would possibly still 
not resemble an FG when merged,
because its neighbourhood is very sparse, i.e., it is 
not embedded in any larger structure, as it is often the case for the
central galaxy in FGs. This is in agreement with its relatively
low bolometric X-ray luminosity of 2.96 $\times$ 10$^{42}$ h$_{70}^{-2}$ 
ergs s$^{-1}$ \citep{Xue2000}.

The environments of nearby CGs have been surveyed by 
\citet{ribeiro98,zabludoff98,carrasco06} among others.  
Spectroscopy of dozens of members in the neighbourhood
of quite a number of groups was obtained, confirming in all cases that
CGs have low velocity dispersions typical of the group regime
(typically 200-300 km s$^{-1}$).  In fact, even for HCG 62, thought to
be one of the most massive CGs in Hickson's catalogue, the
velocity dispersion obtained from 45 members of the system showed that
it is a bonafide group (376 km s$^{-1}$). HCG 62 was suggested by 
\citet{ponman93} as an example of a system that could turn into a FG 
in a few Gygayears, but its velocity dispersion is still much lower
than the value of $\sim$ 600 km s$^{-1}$, typical for
rich FGs. Two other massive nearby CGs
in Hickson's catalogue are HCG 94 and HCG 65.  The first is known
to have a very high bolometric X-ray luminosity of 2.35 $\times$ 10$^{44}$
h$_{70}^{-2}$ \citep{Xue2000} which may, however, be contaminated by
the emission of a nearby cluster. The velocity dispersion obtained from
11 members in this system gives a value of  479 km s$^{-1}$. HCG 65 is the 
center of the cluster Abell 3559. It is in the heart of the Shapley 
supercluster and its location makes it hard to 
disentangle its dynamics and determine its mass. The three most massive 
Hickson CGs known, HCG 62, HCG 65 and HCG 94, are strongly 
early-type dominated, as expected from the velocity dispersion-morphology 
relation observed for CGs. 

The conclusion is then that a typical CG, as observed at z=0, is unlike to
turn into a FG. It is more likely to  merge into an isolated elliptical
galaxy.  For the compact group CG 6, at $z=0.22$ and $\sigma =$703 km
s$^{-1}$,  if we merge the four central galaxies (A, B, C and D), we end up
with a  galaxy with total magnitude \sloani$=$16.31  and \sloang$=$17.80
mag (M$_{i'}$=--23.87 and M$_{g'}$=--22.38, with no k-correction). Using the
color relation for a galaxy at z$=$0.2 from \cite{fuk95},  the magnitude in
r$^{'}$ will be 16.83 or M$_{r}=-23.35$. This magnitude is similar to those
of typical central galaxies in FGs and the velocity dispersion of the
system is typical for the  studied FGs. However, no gap of at least two
magnitudes between the first-ranked  relic and the remaining objects of the
system would be observed because there is at least one other bright galaxy
in the system within half the  virial radius of the group.  We point out
that one other example of a possible massive system, at z=0.39, which may
turn into a FG, has recently  been discovered by \cite{rines07}.
Spectroscopic  studies of CGs at medium redshifts may find many more of
such objects.

\acknowledgments

We acknowledge support from the Brazilian agencies FAPESP (projeto tem\'atico 
01/07342-7), CNPq and CAPES. We thank Laerte Sodr\'e Jr. and Eduardo Cypriano 
for useful discussions in the earlier phase of this project.

\clearpage

\end{document}